\documentclass[11pt,twoside]{article}
\usepackage{asp2014}

\aspSuppressVolSlug
\resetcounters

\begin{document}

\title{A Flexible Expectation Maximization Framework for Fast, Scalable and High-fidelity Multi-frame Astronomical Image Deconvolution}

\author{Yashil~Sukurdeep, Fausto~Navarro and Tam\'{a}s~Budav\'{a}ri}
\affil{Johns Hopkins University, Baltimore, MD, United States}

\paperauthor{Yashil~Sukurdeep}{yashil.sukurdeep@jhu.edu}{0000-0002-4980-4291}{Johns Hopkins University}{Department of Applied Mathematics and Statistics}{Baltimore}{MD}{21218}{United States}
\paperauthor{Fausto~Navarro}{fnavarr3@jhu.edu}{}{Johns Hopkins University}{Department of Applied Mathematics and Statistics}{Baltimore}{MD}{21218}{United States}
\paperauthor{Tam\'{a}s~Budav\'{a}ri}{budavari@jhu.edu}{0000-0002-7034-4621}{Johns Hopkins University}{Department of Applied Mathematics and Statistics}{Baltimore}{MD}{21218}{United States}

\begin{abstract}
We present a computationally efficient expectation-maximization framework for multi-frame image deconvolution and super-resolution. Our method is well adapted for processing large scale imaging data from modern astronomical surveys. Our Tensorflow implementation is flexible, benefits from advanced algorithmic solutions, and allows users to seamlessly leverage Graphical Processing Unit (GPU) acceleration, thus making it viable for use in modern astronomical software pipelines. The testbed for our method is a set of $4$K by $4$K Hyper Suprime-Cam exposures, which are closest in terms of quality to imaging data from the upcoming Rubin Observatory. The preliminary results are extremely promising: our method produces a high-fidelity non-parametric reconstruction of the night sky, from which we recover unprecedented details such as the shape of the spiral arms of galaxies, while also managing to deconvolve stars perfectly into essentially single pixels.
\end{abstract}

\section{Introduction}
\label{sec:intro}
Modern ground-based astronomical surveys, such as the Hyper Suprime-Cam (HSC) Survey~\citep{aihara2018hyper} and the upcoming Legacy Survey of Space and Time (LSST)~\citep{ivezic2019lsst}, will produce wide-field, deep-sky images containing vast amounts of information about the universe. One of the keys to maximizing the insights we gain from this data is \textit{multi-frame deconvolution}. This refers to the process of removing unwanted atmospheric blur from multiple exposures of the same part of the sky in order to produce a single, sharp, high-fidelity image of the night sky. Traditional approaches to address this problem are typically hindered by the varying levels of atmospheric blur across the exposures, their low signal-to-noise ratio, high dynamic range, and the presence of artifacts and obstructions in the telescope's field of view. More recently, streaming methods for multi-frame deconvolution based on the expectation maximization algorithm have been introduced~\citep{harmeling2009online, harmeling2010multiframe, hirsch2011online, lee2017robust, lee2017streaming}. Such frameworks can be used for jointly performing deconvolution and \textit{super-resolution}, which refers to the process of improving the spatial resolution of the reconstruction of the night sky. Yet, these methods face challenges from a computational perspective in the context of modern surveys, largely due to the sheer size of the imaging data, where exposures can contain tens of millions of pixels. We thus build on these approaches and develop a new, flexible expectation-maximization framework for fast, scalable and high-fidelity multi-frame astronomical image deconvolution and super-resolution.

\section{Expectation Maximization Framework}
\label{sec:em_framework}
\begin{figure}[ht]
    \centering   
    \includegraphics[width=0.9\linewidth]{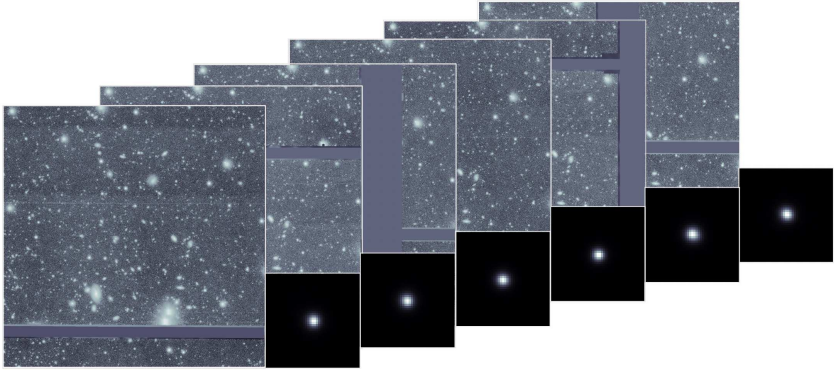}
    \caption{Hyper Suprime-Cam data. We are given a set of $n=33$ exposures $y\!\equiv\!\{y^{(1)}, \dots, y^{(n)}\}$, each of size $4$K by $4$K pixels (left), and corresponding PSFs $f\!\equiv\!\{f^{(1)}, \dots, f^{(n)}\}$ of size $43$ by $43$ pixels (right).}
    \label{fig:HSC_data}
\end{figure}

We start by describing the model for our imaging data. Suppose we are given a set of ground-based exposures of the same part of the sky $y\!\equiv\!\{y^{(1)}, \dots, y^{(n)}\}$, as well as corresponding point-spread functions (PSFs) $f\!\equiv\!\{f^{(1)}, \dots, f^{(n)}\}$, and corresponding variance images $v\!\equiv\!\{v^{(1)}, \dots, v^{(n)}\}$ whose entries $v^{(t)}_{ij}$ represent the variance of photon counts measured at each pixel in each exposure; see Figure~\ref{fig:HSC_data} for an illustration.

We model each observation $y^{(t)}$ as the convolution of a common latent image of the sky, $x$, with the PSF $f^{(t)}$, plus an additive error term $\eta^{(t)}$. Note that the PSFs and error terms can vary from exposure to exposure. The large photon counts in the raw exposures allows us to model the sky-subtracted images as a Gaussian with zero mean and variances $v^{(t)}$. Thus, our model for each pixel value in each exposure $y^{(t)}_{ij}$ is
\begin{equation}
    y^{(t)}_{ij} = \left[f^{(t)} \!* x\right]_{ij} + \eta^{(t)}_{ij}, \quad \textrm{where}\quad \eta^{(t)}_{ij} \sim N\left(0, v^{(t)}_{ij}\right) .
    \label{eq:model_exposures}
\end{equation}

In this setting, an intuitive approach for multi-frame deconvolution, which amounts to finding the unknown latent image $x$, is \textit{maximum likelihood estimation} (MLE). The idea is to estimate the true $x$ by finding an image $\Hat{x}$, called a maximum likelihood estimate, which is most likely to have generated the observed exposures $y$ and PSFs $f$ under our given model. MLE is framed as a constrained optimization problem where one minimizes the (negative) log-likelihood of the latent image $x$. The minimization is typically performed via (stochastic) gradient descent, which converges slowly, and often to undesirable local minima, especially when processing large scale imaging data.

Alternatively, we can use an \textit{expectation-maximization} (EM) approach for finding $\Hat{x}$. Rather than directly minimizing the (negative) log-likelihood function, the idea is to minimize an \textit{auxiliary function} that majorizes the log-likelihood; see~\citet{harmeling2009online} for additional details. For our specific model~\eqref{eq:model_exposures}, the EM approach involves selecting an initial guess $x^{(0)}$ for $\Hat{x}$, and iteratively updating this guess using the \textit{multiplicative update formula} below until some convergence criteria is met:
\begin{equation}
    x^{(k)} = x^{(k-1)} \odot \left( \frac{ \sum_{t=1}^n u^{(t)} \odot \left( {F^{(t)}}^\top y^{(t)} \right) }{ \sum_{t=1}^n u^{(t)} \odot \left( {F^{(t)}}^\top  F^{(t)} x^{(k-1)} \right) } \right) ,
    \label{eq:em_update_l2_loss}
\end{equation}
where $u^{(t)}_{ij} := 1 / v^{(t)}_{ij}$ and where the multiplication and division signs above are defined element-wise. Note that in the expression above, we have also used the fact that convolution is a linear operation to rewrite $f^{(t)} * x = F^{(t)} x$, with $F^{(t)}$ being the linear operator corresponding to the PSF $f^{(t)}$; see~\citet{harmeling2010multiframe} for additional details.

A few aspects of our EM approach for multi-frame astronomical image deconvolution are worth highlighting. Firstly, updating the current iterate $x^{(k-1)}$ using~\eqref{eq:em_update_l2_loss} only involves \textit{element-wise} multiplication of its pixel values, resulting in a procedure that scales well with respect to the size of the imaging data being processed. Secondly, if we start with an initial guess $x^{(0)}$ with \textit{strictly positive} pixel values, the multiplicative update formula guarantees that all of our subsequent iterates (and hence our reconstruction of the night sky $\Hat{x}$) contains non-negative pixel values, which has the effect of speeding up convergence to high-fidelity, physically meaningful images for $\Hat{x}$. Moreover, while our procedure is closely related to the approach of~\citet{harmeling2009online}, a key difference is that we process all frames $y\!\equiv\!\{y^{(1)}, \dots, y^{(n)}\}$ \textit{simultaneously} when updating our estimate for the latent image, while their updates are done in a streaming manner. Consequently, their reconstruction depends on the order in which frames are processed, which can be undesirable. Our approach does not suffer from this shortfall.

Suppose we now wish to also perform super-resolution. More precisely, this involves finding a ``super-resolved" version of the latent image of the sky $x$, denoted by $x^{\Delta}$, whose resolution is higher than that of $x$ by a factor of $\Delta > 1$. We also introduce the \textit{down-sampling operator} $\mathcal{D^r}$, which reduces the resolution of an image by a factor $\Delta > 1$. To perform super-resolution, we consider a modified version of the model from~\eqref{eq:model_exposures}:
\begin{equation}
    \label{eq:model_exposures_super_resolution}
    y^{(t)}_{ij} = \left[ \mathcal{D^r} \left( {h^{(t)}} \!* x^{\Delta} \right) \right]_{ij} + \eta^{(t)}_{ij}, \quad \textrm{where}\quad \eta^{(t)}_{i j} \sim N\left(0, v^{(t)}_{ij}\right),
\end{equation}
where $h\!\equiv\!\{h^{(1)}, \dots, h^{(n)}\}$ are ``super-resolved" versions of the PSFs $f\!\equiv\!\{f^{(1)}, \dots, f^{(n)}\}$. We self-consistently compute each $h^{(t)}$ as follows:
\begin{equation}
    \label{eq:psfs_super_resolution}
    h^{(t)} = \underset{h \geq 0}{\operatorname{argmin}} \quad \sum_{i,j} \left( \left[ \mathcal{D^r}(h * g) \right]_{ij} - f^{(t)}_{ij} \right)^2 + \lambda \left( \nabla h \right)_{ij}^2 ,
\end{equation}
where $g$ is a fixed Gaussian PSF with width $\Delta > 1$, $\lambda > 0$ is a balancing hyper-parameter, and $(\nabla h)_{ij}$ denotes the gradient of PSF $h$ at pixel $(i,j)$. This computation yields PSFs $h^{(t)}$ with a resolution $\Delta$ times higher than $f^{(t)}$, where the down-sampled version of $h^{(t)}$ is similar to $f^{(t)}$ up to a convolution with a small Gaussian PSF $g$. To perform multi-frame deconvolution and super resolution jointly, we simply use the super-resolved PSFs $h$ and up-sampled versions of the exposures $y^{\Delta}$ (rather than the original PSFs $f$ and exposures $y$) and apply the multiplicative update procedure in~\eqref{eq:em_update_l2_loss}.

\begin{figure}[ht]
    \centering
    \includegraphics[width=0.321\linewidth]{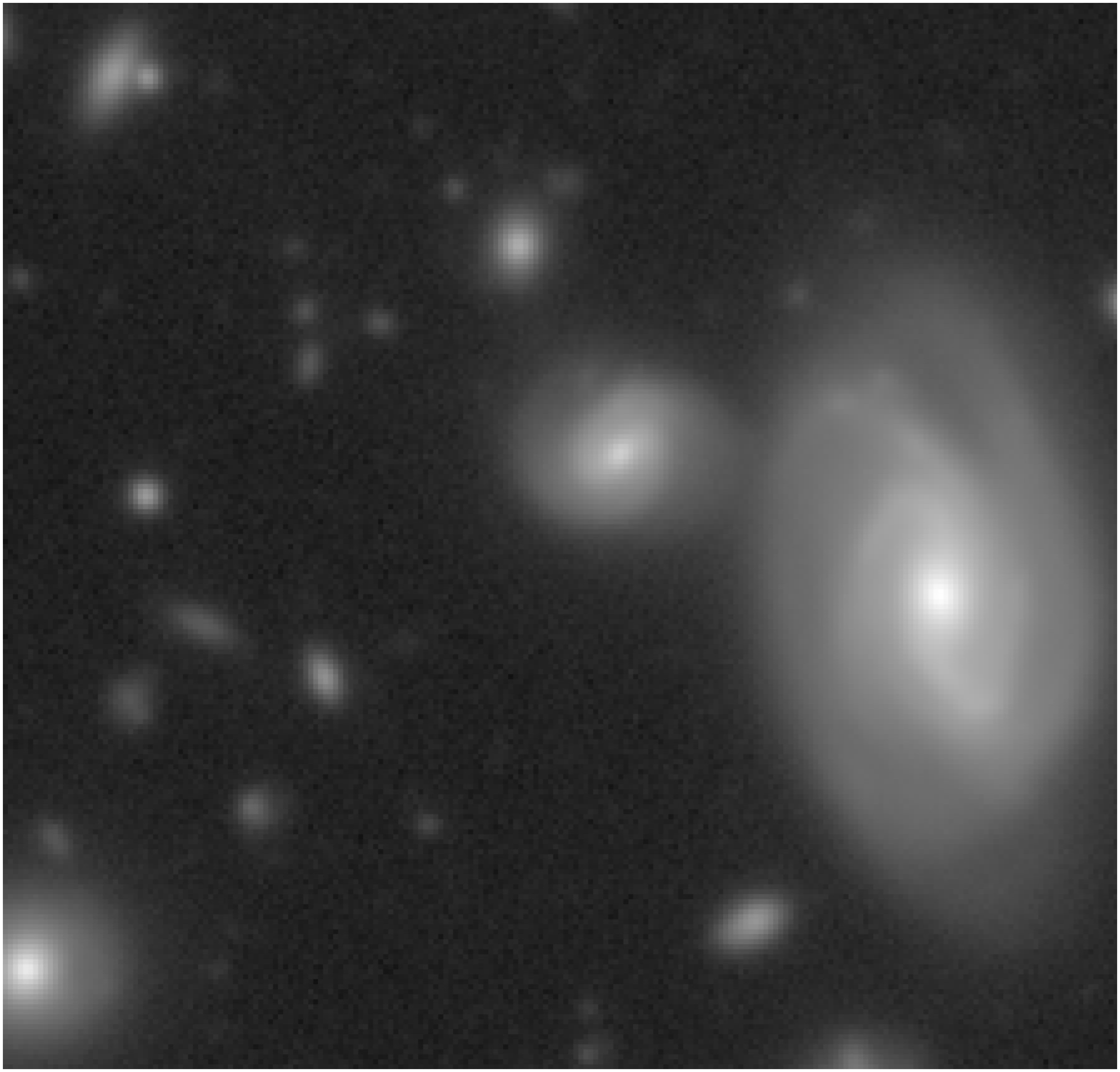}
    \includegraphics[width=0.323\linewidth]{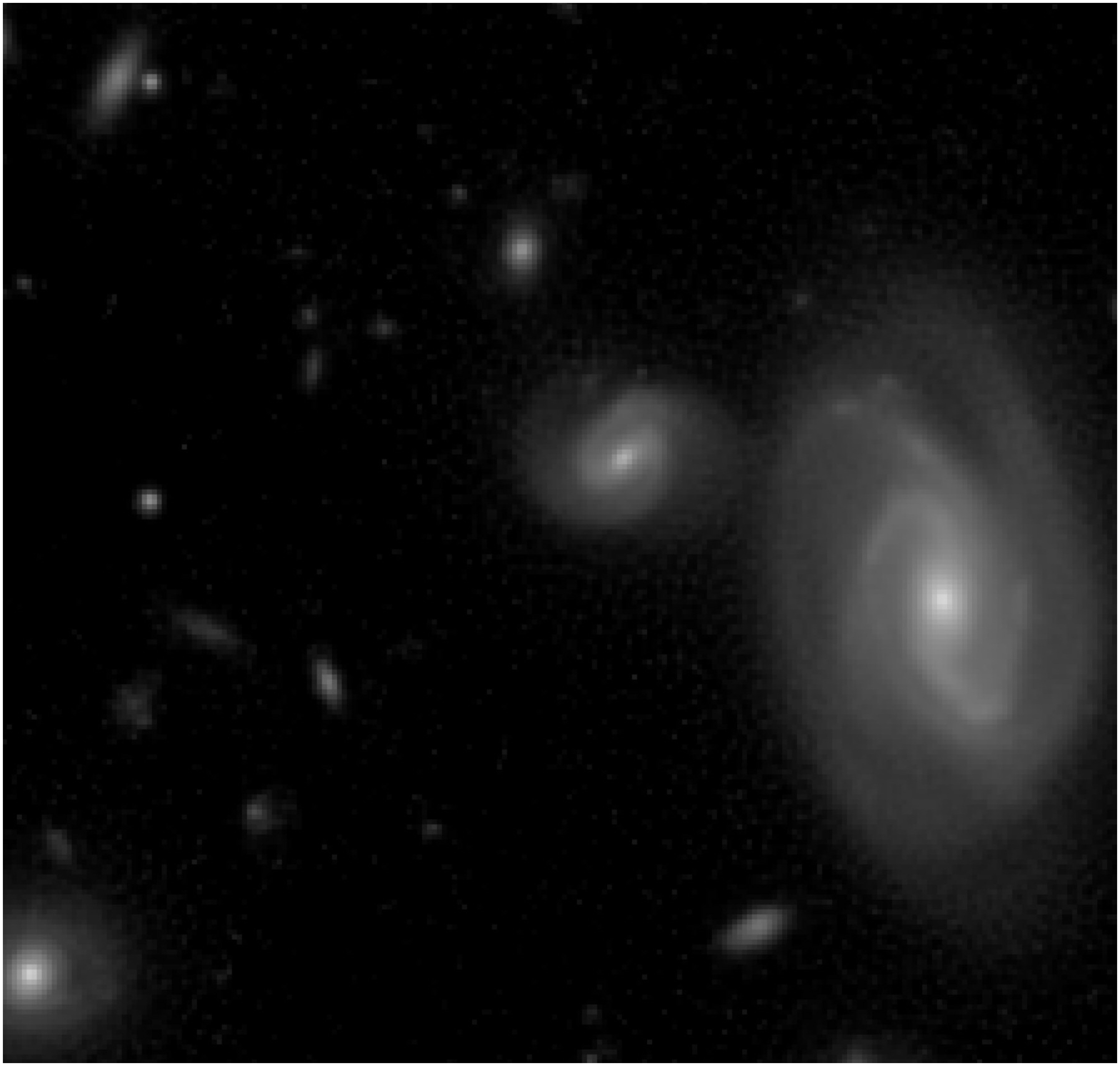}
    \includegraphics[width=0.32\linewidth]{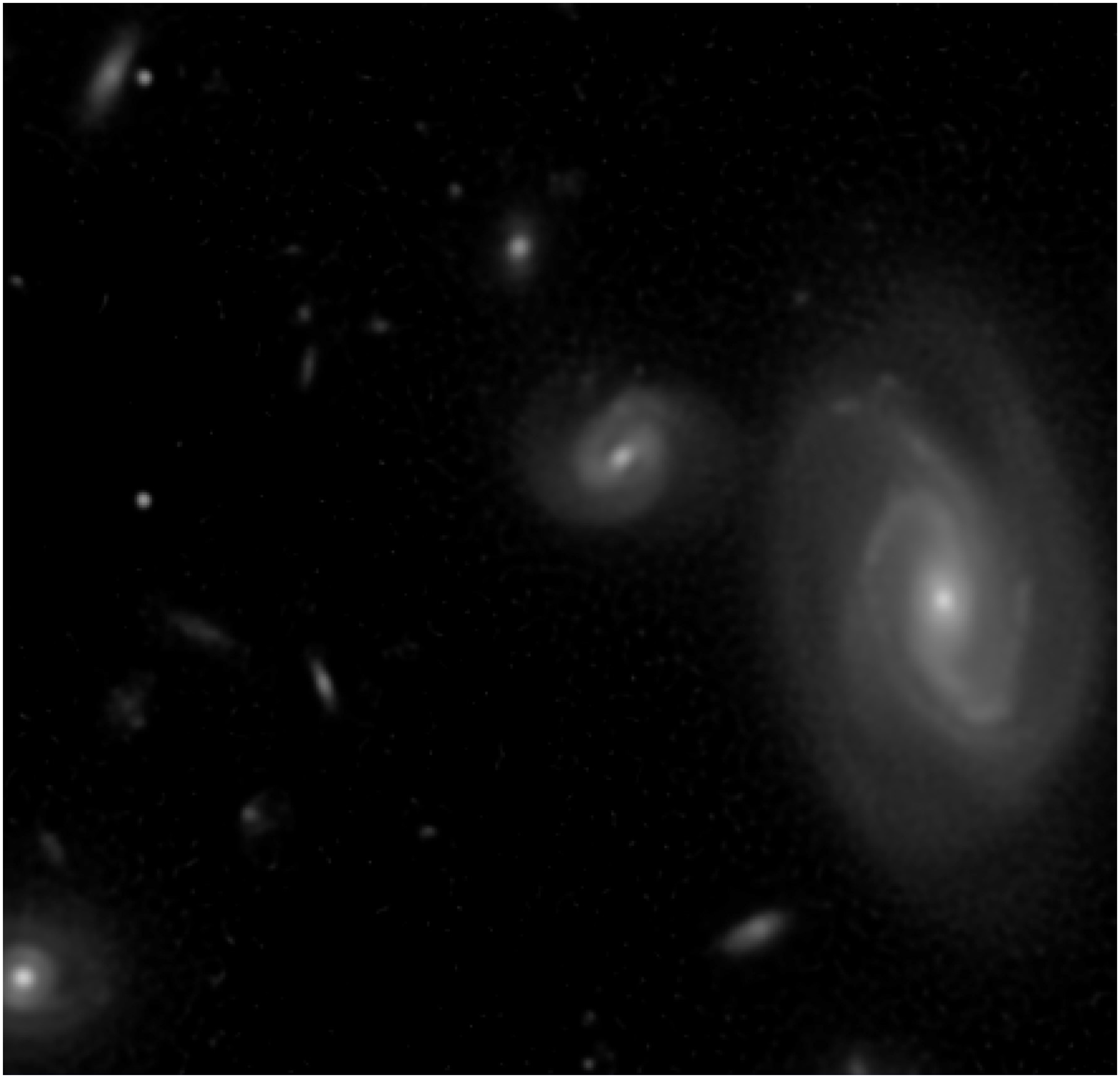}
    \includegraphics[width=0.321\linewidth]{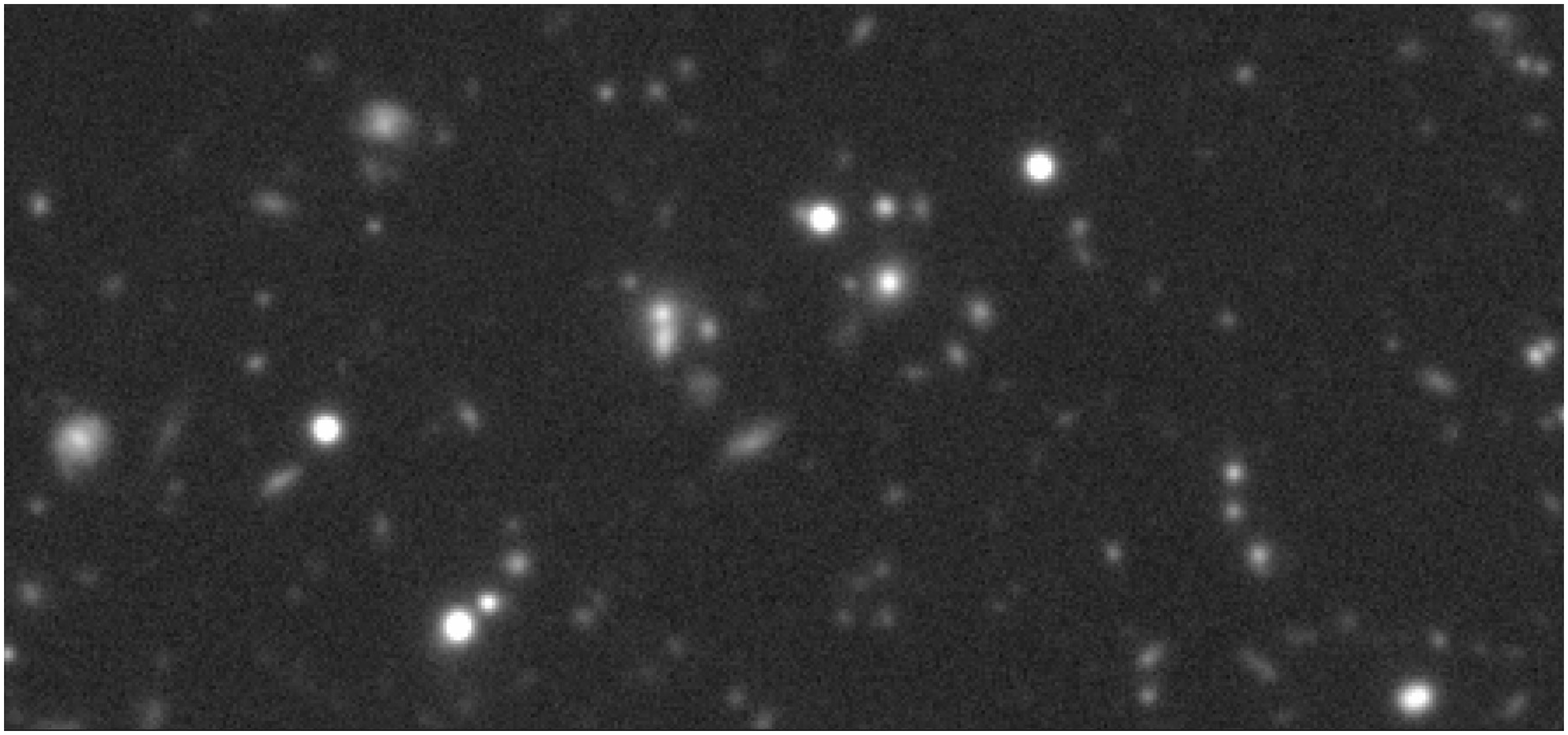}
    \includegraphics[width=0.323\linewidth]{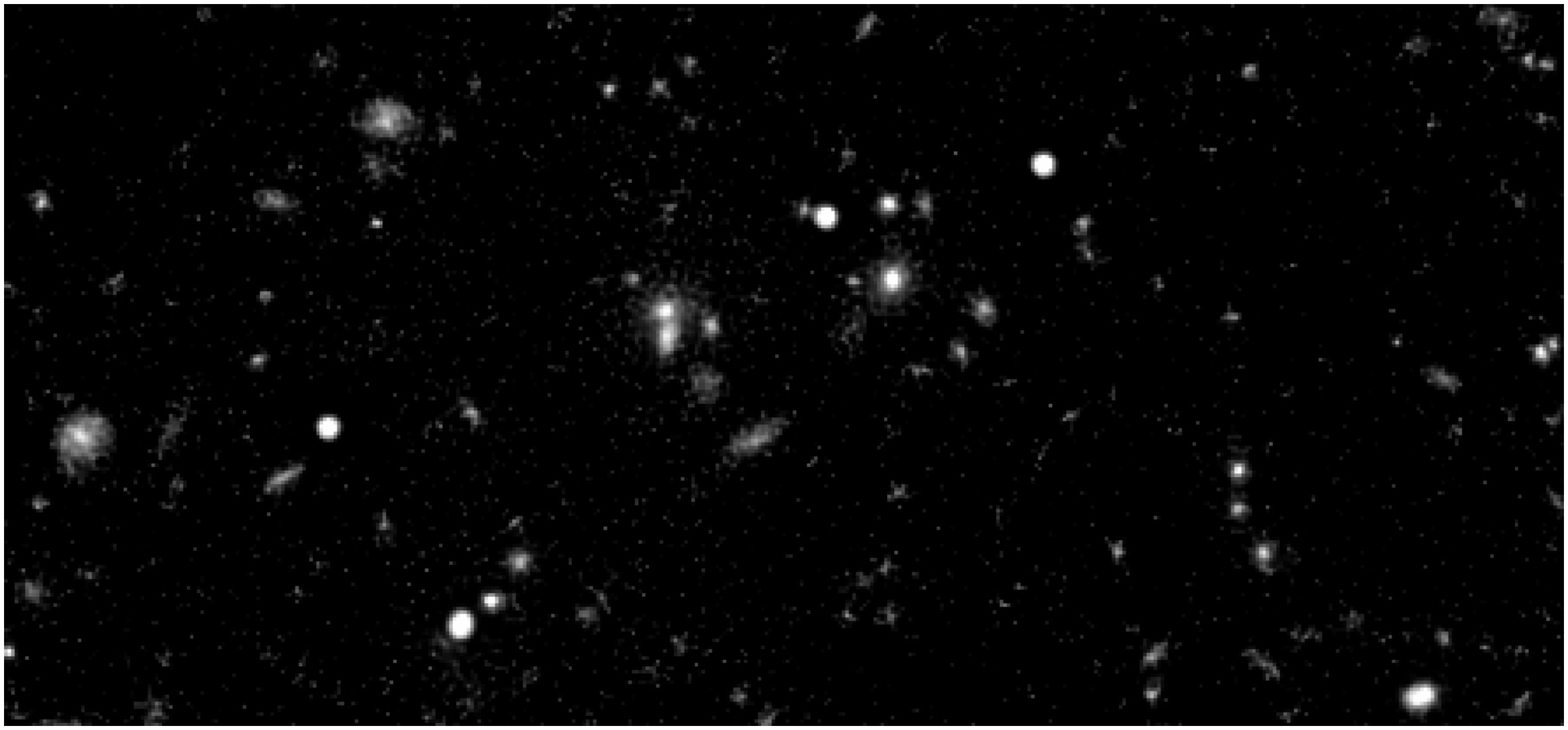}
    \includegraphics[width=0.32\linewidth]{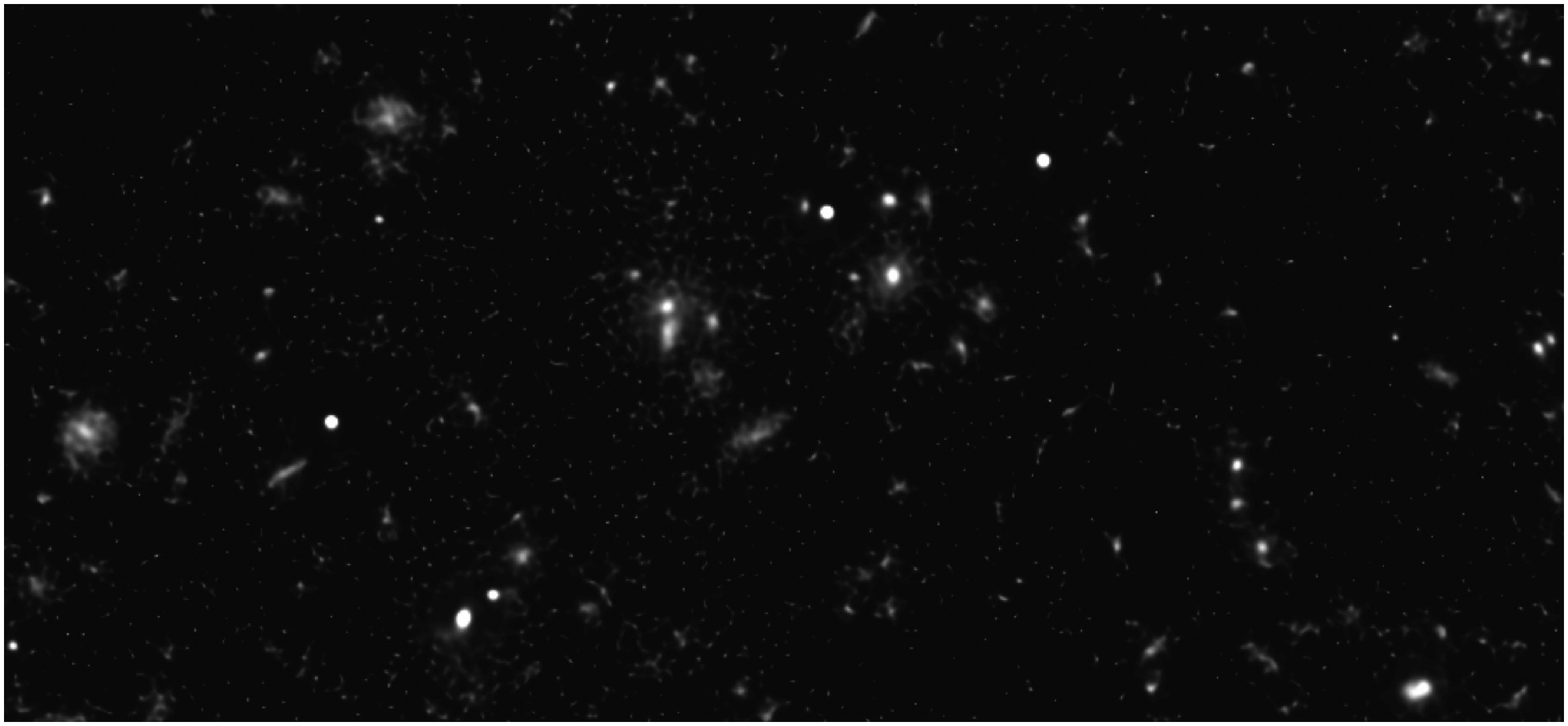}
    \caption{Comparison: Selected cutouts from a ``sample mean" co-add (left) vs. a restored image $\Hat{x}$ from our EM approach without super-resolution (middle), and with super-resolution using $\Delta = 2$ (right). Our method de-blurs a wide array of sources, such as spiral arm and elliptical galaxies, and stars of varying sizes and shapes (top row), as well as small, faint sources (bottom row). The reconstructions contain none of the usual unwanted artifacts (e.g ringing, speckles, noise in the sky background), producing high-fidelity images where e.g., the pixels are non-negative, the sky-background has zero pixel values, and the number and relative sizes, shapes and fluxes of the sources is preserved. Overall, the method produces a physically meaningful restored image of the night sky which is suitable for photometry, especially when super-resolution is used.}
    \label{fig:comparison_results_coadd}
\end{figure}

We implemented our algorithm in \texttt{Tensorflow}, leveraging its built-in implementation of the Adam algorithm to obtain the super-resolved PSFs by solving~\eqref{eq:psfs_super_resolution}, and using Graphical Processing Unit (GPU) acceleration to perform our EM updates. We tested our method on a set of $33$ exposures from the HSC telescope, with results in Figure~\ref{fig:comparison_results_coadd}. Performance wise, it took $60$ seconds on average to perform multi-frame deconvolution and super-resolution on this dataset, \textit{with} GPU-acceleration.

As future work, the authors will carry photometric and statistical tests on the resulting reconstructions. Furthermore, we plan to incorporate improved sky-background subtraction in the framework, as well as robust statistics in the model by using distributions with heavier tails for the noise terms in~\eqref{eq:model_exposures}, which should in theory improve the reconstructions in the presence of extreme outliers (e.g. satellite trails) in the exposures.

\bibliography{C27}

\end{document}